\documentclass[12pt]{elsart}
\usepackage{amsmath,amssymb}
\usepackage{graphicx,psfrag,here,epsfig}
\usepackage{pstricks}
\usepackage{pst-node}
\usepackage{epsf}
\usepackage{citesort}
\allowdisplaybreaks[1]

\begin{document}
%%%%%%%%%%%%%%%%%%%%%%%%%%%%%%%%%%%%%%%%%%%%%%%%%%%%%%%%%%%%%%%%%%%%%%%%%%%%%
\newcommand{\co}{\; \; ,}
\def\words#1{\mbox{\small{\,#1}}}
\def\bea{\begin{eqnarray}}
\def\eea{\end{eqnarray}}
\def\eq{\begin{eqnarray}}
\def\en{\end{eqnarray}}
\def\be{\begin{equation}}
\def\ee{\end{equation}}
\newcommand{\ed}{\end{document}}
\newcommand{\rr}{\mbox{\boldmath $r$}}
\newcommand{\rrn}{\mbox{$r$}}
\newcommand{\rp}{\mbox{\boldmath $p$}}

\newcommand{\nnnl}{\nonumber\\}
\newcommand{\fs}{\, . \,}
\def\query#1{\marginpar{\begin{flushleft}\footnotesize#1\end{flushleft}}}%

\runauthor{Gay Ducati, Machado and Machado}
\renewcommand{\theequation}{\arabic{equation}}
\begin{frontmatter}

\title{\Large\bf Small-$x$ neutrino-hadron structure functions within the QCD color dipole picture}

\author[UFRGS]{M.B.~Gay Ducati,}
\author[UFRGS]{M.M.~Machado,}
\author[UNIPAMPA]{M.V.T.~Machado,}

\address[UFRGS]{High Energy Physics Phenomenology Group, GFPAE,  IF-UFRGS \\
Caixa Postal 15051, CEP 91501-970, Porto Alegre, RS, Brazil}
\address[UNIPAMPA]{Centro de Ci\^encias Exatas e Tecnol\'ogicas, Universidade Federal do Pampa \\
Campus de Bag\'e, Rua Carlos Barbosa. CEP 96400-970. Bag\'e, RS, Brazil}

\begin{abstract}
  We present an exploratory QCD analysis of the neutrino structure functions in charged current DIS using the color dipole formalism. The dipole cross sections are taken from recent phenomenological/theoretical studies in deep inelastic inclusive production. The theoretical predictions are compared to the available experimental results in the small-$x$ region, which has never been considered so far.
\end{abstract}


\end{frontmatter}

\section{Introduction}

% Motivation
% ------------------------------
The interaction of  high energy neutrinos on hadron targets are an outstanding probe to test Quantum Chromodynamics (QCD) and understanding the parton properties of hadron structure. The several combinations of neutrino and anti-neutrino scattering data can be used to determine the structure functions, which constrain the valence, sea and gluon parton distributions in the nucleons/nuclei. The neutrino structure functions are needed for computing the total neutrino-hadron cross section, which plays an important role in high energy cosmic rays studies and in astroparticle physics \cite{STASTO}. The differential cross section for the neutrino-nucleon charged current process $\nu_l\,(\bar{\nu}_l)+N \rightarrow l^-\,(l^+)+X$, in terms of the Lorentz invariant structure functions $F_2^{\nu N}$, $2xF_1^{\nu N}$ and $xF_3^{\nu N}$ are \cite{Leader_Predazzi},
\begin{eqnarray}
\frac{d\sigma^{\nu ,\bar{\nu}}}{dx \,dy} =  \frac{G_F^2 \,m_N \,E_{\nu}}{\pi }\left[ \left( 1-y-\frac{m_Nxy}{2E_{\nu}} \right) F_2+ \frac{y^2}{2} \,2xF_1 \pm   y\left( 1-\frac{y}{2} \right) xF_3\right]\,,\nonumber
\end{eqnarray}
where $G_F$ is the weak Fermi coupling constant, $m_N$ is the nucleon mass, $E_{\nu}$ is the incident neutrino energy, $Q^2$ is the square of the four-momentum transfer to the nucleon. The variable $y=E_{had}/E_{\nu}$ is the fractional energy transferred to the hadronic vertex with $E_{had}$ being the measured hadronic energy, and $x=Q^2/2m_NE_{\nu}y$ is the Bjorken scaling variable (fractional momentum carried by the struck quark).

Similarly to the charged-lepton DIS, the deep inelastic neutrino scattering is also used to investigate the structure of nucleons and nuclei. In the leading order quark-parton model (the QCD collinear approach), the structure function $F_2$ is the singlet distribution, $F_2^{\nu N}\propto xq^S=x\sum(q+\bar{q})$, the sum of momentum densities of all interacting quarks constituents, and $xF_3$ is the non-singlet distribution, $xF_3^{\nu N}\propto xq^{NS}=x\sum(q-\bar{q})=xu_V+xd_V$, the valence quark momentum density. These relations are further modified by higher-order QCD corrections.  Currently, the theoretical description of experimental data on neutrino DIS is reasonable (for a very recent investigation, see Ref. \cite{duancg}). The main theory uncertainties are the role played by nuclear shadowing in contrast with lepton-charged DIS and a correct understanding of the low $Q^2$ limit. The first uncertainty can be better addressed with the future precise data from MINER$\nu$A \cite{minerva}. However, nuclear effects are taken into account by using the nuclear ratios $R=F_2^A/AF_2^p$ extracted from lepton-nucleus DIS, which it could be different for the neutrino-nucleus case. The low-$Q^2$ region can not be addressed within the pQCD quark-parton model as a hard momentum scale $Q_0^2\geq 1-2$ GeV$^2$ is required in order to perform perturbative expansion.

In this work we present a determination of the small-$x$ structure functions for neutrino-nucleus within the color dipole formalism \cite{DIPOLEPIC}. This approach allows for a simple implementation of shadowing corrections \cite{ARMESTO} in neutrino-nuclei interactions. This paper is organized as follows. In Sec. 2, the structure function $F_2^{\nu N}$ is investigated within the color dipole picture at small-$x$ region, employing recent phenomenological parton saturation models.  It is shown that small-$x$ data exhibit geometric scaling property, which  has important consequences for ultra-high energy neutrino phenomenology. In Sec. 3, the structure function $xF_3$ and the quantity $\Delta x F_3$ are addressed. The latter one provides a determination of the strange-sea parton distribution through charm production in charged-current neutrino DIS. Finally, we also analyze the nuclear ratios $R_2 = F_2^{\nu A}/AF_2^{\nu N}$ and $R_2 = xF_3^{\nu A}/AxF_3^{\nu N}$. In the last section we present comments and conclusions.

\section{Neutrino Structure Function $F_2^{\nu N}(x,Q^2)$}

We focus here in the high energy regime, which one translates into small-$x$ kinematical region. At this domain a quite successful framework to describe QCD interactions is provided by the color dipole formalism \cite{DIPOLEPIC}, which allows an all twist computation  (in contrast with the usual leading twist approximation) of the structure functions. The physical picture of the interaction is the deep inelastic scattering viewed as the result of the interaction of a color $q \bar{q}$ color dipole, in which the gauge boson fluctuates into, with the nucleon target. The interaction is modeled via the dipole-target cross section, whereas the boson fluctuation in a color dipole is given by the corresponding wave function. The charged current (CC)  DIS structure functions \cite{BGNPZ1,BGNPZ2,KUTAK} are related to the cross section for scattering of transversely and longitudinally polarized $W^{\pm}$ bosons. That is,
\begin{eqnarray}
F_{T,L}^{\mathrm{CC}}\,(x,Q^2) = \frac{Q^2}{4\,\pi^2}\, \int d^2 \rr \,\int_0^1 dz \,
| \psi^{W^{\pm}}_{T,L}\,(z,\,\rr,\,Q^2)|^2\,\sigma_{dip}\,(x,\,\rr)\,,
\label{FSDIP}
\end{eqnarray}
where  $\rr$ denotes the transverse size of the color dipole, $z$ the
longitudinal momentum fraction carried by a quark and  $\psi^{W}_{T,L}$ are  the light-cone wavefunctions for (virtual) charged gauge bosons with transverse or longitudinal polarizations. The small-$x$ neutrino structure function $F_2^{\nu N}$ is computed from expressions above taking $F_2=F_T+F_L$. Explicit expressions for the wave functions squared  can be found at Refs. \cite{BGNPZ1,BGNPZ2,KUTAK}. In what follows we consider four quark flavors ($u,d,s,c$) with masses $m_f$. The color dipoles  contributing  to Cabibbo favored transitions are  $ u \bar d \, (d \bar u)$,  $ c \bar s \,(s \bar c)$ for CC interactions.

The dipole hadron cross section $\sigma_{dip}$  contains all
   information about the target and the strong interaction physics. In the present study, we consider analytical expressions for the dipole cross section, with particular interest for those ones presenting scaling behavior. Namely, one has $\sigma_{dip} \propto (\rr^2Q^2_{\mathrm{sat}})^\gamma$ for dipole sizes $\rr^2 \approx 1/Q^2_{\mathrm{sat}}$ and where $(1-\gamma )$ is the effective anomalous dimension. The so-called saturation scale $Q_{\mathrm{sat}} \propto x^{\lambda/2}$ defines the onset of the parton saturation effects. In what follows one takes the phenomenological parameterizations: (a) Golec-Biernat-W\"{u}sthoff model (GBW) \cite{GBW} and  (b) Itakura-Iancu-Munier (IIM) model \cite{IIM}. Both models are able to describe experimental data on inclusive and diffractive deep inelastic $ep$ scattering at small-$x$. We quote the original papers for details on the parameterizations and determination of their phenomenological parameters. We call attention to Ref. \cite{Jamal}, where the implications of saturation in ultrahigh energy neutrino cross section has been  first discussed within the color dipole picture. We will show that the results are not strongly sensitive to a particular choice of model. Here, we use an effective light quark mass, $m_f=0.14$
GeV and the charm mass is set to be $m_c=1.5$ GeV. In addition, as the color dipole models are suitable in the region below $x=0.01$ and the
large $x$ limit still needs a consistent  treatment,  we supplement
 the dipole cross sections with a threshold factor
$(1-x)^{n_{\mathrm{thres}}}$ ($n_{\mathrm{thres}}=5\,(7)$ for number of flavors $n_f=3\,(4)$).

The extension of the approach to consider nuclei targets we take the Glauber-Gribov picture \cite{ARMESTO}, without any new parameter. In this
approach, the nuclear version is obtained replacing the
dipole-nucleon cross section  by the
nuclear one,
\begin{eqnarray}
\sigma_{dip}^{\mathrm{nucleus}} (x, \,\rr^2;\, A)  = 2\,\int d^2b \,
\left\{\, 1- \exp \left[-\frac{1}{2}\,T_A(b)\,\sigma_{dip}^{\mathrm{nucleon}} (x, \,\rr^2)  \right] \, \right\}\,,
\label{sigmanuc}
\end{eqnarray}
where $b$ is the impact parameter of the center of the dipole
relative to the center of the nucleus and the integrand gives the
total dipole-nucleus cross section for a  fixed impact parameter.
The nuclear profile function is labeled by $T_A(b)$ \cite{devries}.

%%%%%%%%%%%%%%%%%%%%%%%%%%%%%%%%%%%%%%%%%%%%%%%%%%%%%%%%%%%%%%%%%%%%

\begin{figure}[t]
\centerline{\includegraphics[scale=0.45]{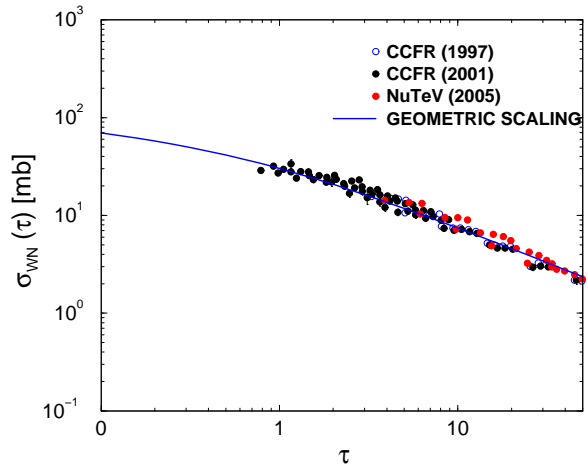}}
\caption{The scaling behavior of $\sigma_{WN}(x,Q^2)$ as a function of scaling variable $\tau$ (see text).}
\label{fig:1}
\end{figure}
%%%%%%%%%%%%%%%%%%%%%%%%%%%%%%%%%%%%%%%%%%%%%%%%%%%%%%%%%%%%%%%%%%%%%

A basic property of
the saturation physics is the geometric scaling. It means that  the total $\gamma^* p$
cross section at large energies is not a function of the two
independent variables $x$ and $Q$, but is rather a function of the
single variable $\tau_p = Q^2/Q_{\mathrm{sat}}^2(x)$ as shown
in Ref. \cite{SGK}. That is, $\sigma_{\gamma^*p}(x,Q^2)=\sigma_{\gamma^*p}(\tau_p)$. In Refs. \cite{IIML,MT} it was shown that the geometric scaling observed in experimental data can be understood theoretically in the context of non-linear QCD evolution with fixed and running coupling. Recently, the high energy $l^{\pm}p$,  $pA$ and $AA$ collisions have been related through geometric scaling \cite{Armesto_scal}. Within the color dipole picture and making use of a rescaling of the impact parameter of the $\gamma^*h$ cross section in terms of hadronic target radius $R_h$, the nuclear dependence of the $\gamma^*A$ cross section is absorbed in the $A$-dependence of the saturation scale via geometric scaling. The relation reads as $\sigma^{\gamma^*A}_{tot}(\tau_A)  =  \kappa_A\,\sigma^{\gamma^*p}_{tot}\,(\tau_A)$, where $\kappa_A = (R_A/R_p)^2$. The nuclear saturation scale was assumed to rise with the quotient of the transverse parton densities to the power $\Delta $ and $R_A$ is the nuclear radius, $Q_{\mathrm{sat},A}^2=(A/\kappa_A)^{\Delta}\,Q_{\mathrm{sat},p}^2$.  The functional shape for the photoabsortion cross section, $\sigma_{\gamma^*p}\,(\tau_p)$, has been considered based on theoretical studies \cite{Armesto_scal}.

The geometric scaling property has direct consequences on the computation of small-$x$ neutrino structure functions.  It has been shown in Ref. \cite{MPRD1} that the charged (CC)  and neutral (NC) current structure functions are described by the same mathematical expressions as the proton structure function up to a different coupling of the electroweak bosons. Thus, using such a property one has for the neutrino-nuclei  collisions \cite{MPRD1},
\begin{eqnarray}
\sigma_{tot}^{W^{\pm}A}\,(x,\,Q^2;\,A)  = \bar{\sigma}_0\,\left(\frac{n_f\kappa_A}{\alpha_{\mathrm{em}}\hat{e}_f^2}\right)\,
  \left[ \gamma_E + \Gamma\left(0,\frac{a}{\tau_A^b} \right) +
         \ln \left(\frac{a}{\tau_A^b}\right) \right],
	 \label{sigboscc}
\end{eqnarray}
where $\alpha_{\mathrm{em}}$ is the QED constant coupling, $\hat{e}_f$ is the sum of electric charge of the quarks of flavor $f$. The quantity $\gamma_E$ is the Euler constant and $\Gamma\left(0,\beta\right)$
the incomplete Gamma function. The parameters $a$ and $b$  were obtained from a fit to the small-$x$ $ep$ DESY-HERA data, with the overall  normalization fixed by $\bar\sigma_0=40.56$ $\mu$b \cite{Armesto_scal}.

%%%%%%%%%%%%%%%%%%%%%%%%%%%%%%%%%%%%%%%%%%%%%%%%%%%%%%%%%%%%%%%%%%%%

\begin{figure}[t]
\centerline{\includegraphics[scale=0.55]{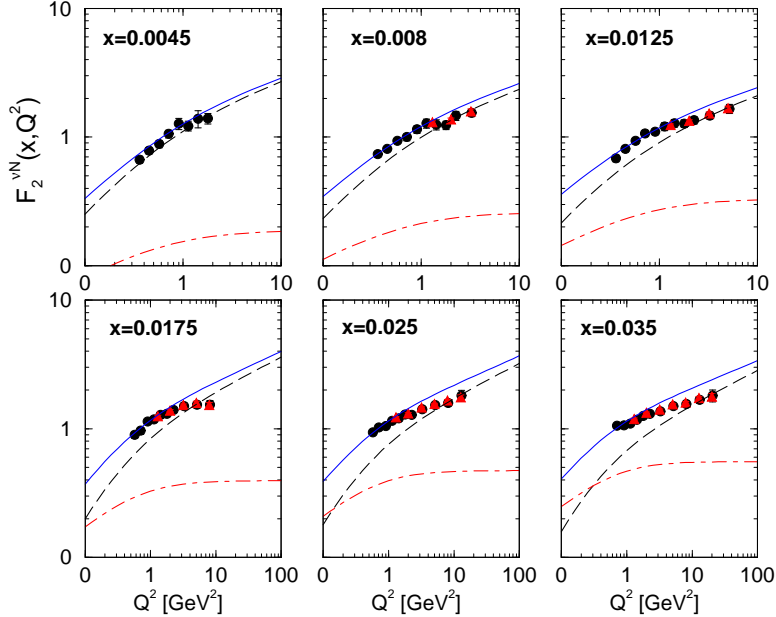}}
\caption{The structure function $F_2^{\nu N}(x,Q^2)$ as a function of boson virtuality. The long-dashed curve corresponds  to the geometric scaling result for the neutrino-nucleus interactions, the dot-dashed curve to the valence content and the solid one is the sum of both contributions. Experimental data from CCFR Collaboration \cite{CCFR,Fleming}.}
\label{fig:2}
\end{figure}
%%%%%%%%%%%%%%%%%%%%%%%%%%%%%%%%%%%%%%%%%%%%%%%%%%%%%%%%%%%%%%%%%%%%%

It is important to investigate whether this geometric scaling pattern is exhibited by the small-$x$ neutrino deep inelastic data. In order to do so, we take the datasets for the structure function $F_2$ with the kinematical cut $x\leq 0.035$ and all $Q^2$ \cite{CCFR,Fleming,Tzanov}. These experiments measure the differential cross sections for deep inelastic $\nu_{\mu}$-Fe and $\bar{\nu}_{\mu}$-Fe  scattering and the structure functions are then extracted. In Fig. \ref{fig:1}, we plot the quantity $\sigma_{WN}=\frac{4\pi^2}{Q^2}F_2^{\nu N}$ as a function of the scaling variable $\tau = Q^2/Q_{\mathrm{sat}}^2(x)$. The saturation scale squared reads as $Q_{\mathrm{sat}}^2(x)=(3\cdot 10^{-4}/x)^{0.288}$ \cite{GBW}. The geometric scaling behavior is present with a smooth spread around the scaling curve. The scaling curve (solid line) is obtained using Eq. (\ref{sigboscc}). This result is useful and can be investigated in more detail using precise measurements in future experiments. In the present kinematical window, the color dipole formalism (and geometric scaling property) is in the limit of its validity. However, the present result shows that it gives a reasonable phenomenological description of the limit case.

Now, we will compare the color dipole prediction against the structure function $F_2^{\nu N}=\frac{Q^2}{4\pi}\sigma_{WN}$. This is presented in Fig. \ref{fig:2}. We use the experimental datasets of the CCFR Collaboration \cite{CCFR,Fleming}, where filled circles correspond to points in Ref. \cite{Fleming} and triangles up correspond to points in Ref. \cite{CCFR}. The long-dashed curve is obtained using scaling expression Eq. {\ref{sigboscc}) for the boson-nucleus cross section. Nuclear effects are taken into account through the nuclear saturation scale. The calculation produces a suitable description at small-$x$, despite the data points lying at the expected validity region of the color dipole approach, $x\leq 10^{-2}$. For completeness, we have added the valence content to $F_2$. To this aim, the following parameterisation has been considered \cite{Reno},
\begin{eqnarray}
F_2^{val} & = & B_{\nu}\,x^{1-\alpha_R}(1-x)^{n(Q^2)}\,\left( \frac{Q^2}{Q^2+b}\right)^{\alpha_R}\,[1+f_{\nu}(1-x)]\,,
\end{eqnarray}
which it is a Regge inspired model. Above, $n(Q^2) = \frac{3}{2}\left(1+\frac{Q^2}{Q^2+c}\right)$. The parameters are taken from Ref. \cite{Reno}, with $B_{\nu}= 2.695$, $f_{\nu}=0.595$, $\alpha_R=0.425$, $c=3.5489$ GeV$^2$ and $b=0.6452$ GeV$^2$.
The valence contribution is represented by the dot-dashed curve, which is subleading at small-$x$, but improves the overall description. The total contribution (valence + sea color dipole) is given by the solid curve.  It is worth to mention that we have not tunned the original parameters of the scaling model and of the valence parametrization.

A short comment is in order here. We found that the complete numerical calculation for the structure function $F_2$, using Eqs. (\ref{FSDIP})  and nuclear corrections from Eq. (\ref{sigmanuc})  produces results compatible with the geometric scaling calculation above within a few percents. This fact corroborates the advantage of using the fast scaling parametrization, which can be promptly computed for any nuclei. This is possible because the parameters of the scaling curve  suitably absorb the details of the charm contribution, which turn out the complete calculation somewhat more involved. In the next sections, we will use the full color dipole calculation as the structure function $xF_3$ is strongly sensitive to the charm quark contribution.

\section{The quantities  $xF_3^{\nu N}$,  $\Delta xF_3^{\nu N}$ and the nuclear ratios}

%%%%%%%%%%%%%%%%%%%%%%%%%%%%%%%%%%%%%%%%%%%%%%%%%%%%%%%%%%%%%%%%%%%%

\begin{figure}[t]
\centerline{\includegraphics[scale=0.5]{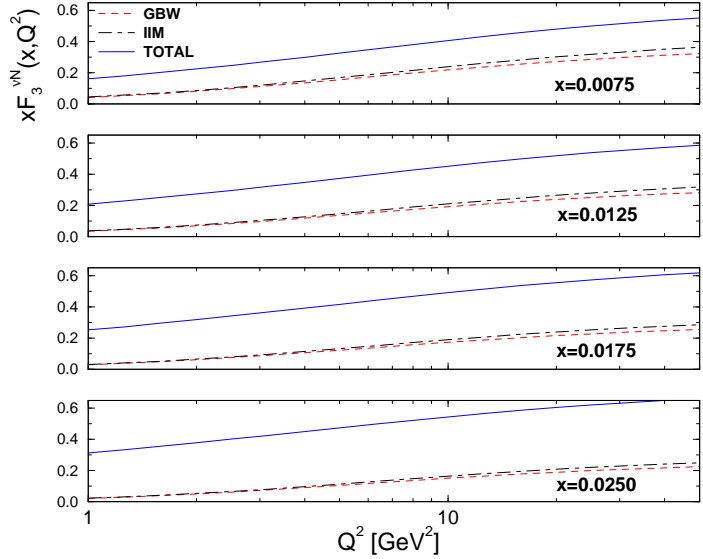}}
\caption{The structure function $xF_3^{\nu N}(x,Q^2)$ as a function of boson virtuality (see text).}
\label{fig:3}
\end{figure}
%%%%%%%%%%%%%%%%%%%%%%%%%%%%%%%%%%%%%%%%%%%%%%%%%%%%%%%%%%%%%%%%%%%%%

Lets now compute the structure functions $xF_3^{\nu N}$ within the color dipole formalism. We concentrate on the
interaction of the $c\bar s$ color dipole of size $\rr$
 with the target  hadron which is described by the beam-
 and flavor-independent color dipole cross section
$\sigma_{dip}$. In the infinitum momentum frame, this is equivalent to the $W^{\pm}$-gluon fusion process, $W^{\pm}+g\rightarrow  c\bar{s} \,(\bar{c}s)$.
The analysis for charged current DIS has been addressed in Refs. \cite{ZOLLER1,ZOLLER2}, where the left-right asymmetry  of diffractive
interactions of electroweak bosons of different helicity is discussed. There, the relevant light-cone wavefunctions have been evaluated. The contribution of excitation of open
charm/strangeness to the hadron absorption cross section for 
left-handed ($L$)  and right-handed ($R$) $W$-boson of virtuality $Q^2$,
is given by \cite{DIPOLEPIC},
\begin{eqnarray}
\sigma_{L,\,R}\,(x,Q^{2})
=\int d^{2}{\rr} \int_0^1 dz \sum_{\lambda_1,\lambda_2}
|\Psi_{L,\,R}^{\lambda_1,\lambda_2}(z,\rr,Q^2)|^{2}
\,\sigma_{dip}\,(x,\rr )\,,
\label{eq:FACTOR}
\end{eqnarray}
where $\Psi_{L,\,R}^{\lambda_1,\lambda_2}(z,\rr,Q^2)$
 is the light-cone wavefunction of
the $c\bar{s}$ state with the $c$ quark
carrying fraction $z$ of the $W^+$ light-cone momentum and
$\bar s$ with momentum fraction $1-z$. The $c$- and $\bar s$-quark
helicities are  $\lambda_1=\pm 1/2$ and  $\lambda_2=\pm 1/2$, respectively.
The diagonal elements of  density matrix are given by,
\begin{eqnarray}
\sum_{\lambda_1,\lambda_2}\Psi_{L}^{\lambda_1,\lambda_2}
\left(\Psi_{L}^{\lambda_1,\lambda_2}\right)^* & = & \frac{4\,N_c}{(2\pi)^2}\,z^2\,\left[m_{\bar{q}}^2\,K_0^2(\varepsilon r)+ \varepsilon^2\,K_1^2(\varepsilon r)\right]\,,\label{eq:RHOR}\\
\sum_{\lambda_1,\lambda_2}\Psi_{R}^{\lambda_1,\lambda_2}
\left(\Psi_{R}^{\lambda_1,\lambda_2}\right)^* & = & \frac{4\,N_c}{(2\pi)^2}\,(1-z)^2\,\left[m_{q}^2\,K_0^2(\varepsilon r)+ \varepsilon^2\,K_1^2(\varepsilon r)\right]\,.
\label{eq:RHO}
\end{eqnarray}

In the expressions given by  Eqs. (\ref{eq:RHOR})-(\ref{eq:RHO}), one uses the notation $\varepsilon^2=z(1-z)Q^2+(1-z)m_q^2+zm_{\bar{q}}^2$, where the quark
and antiquark masses are $m_q$ and $m_{\bar{q}}$, respectively.
The corresponding expressions for $ W^{-}$ boson are obtained by replacing
$m_q\leftrightarrow m_{\bar{q}} $. It should be noticed the strong left-right asymmetry referred above.

The structure function of deep inelastic neutrino-nucleon  $xF_3$ can be defined in terms of $\sigma_{R}$ and $\sigma_{L}$ of Eq.~(\ref{eq:FACTOR}) in the following usual way,
\begin{eqnarray}
xF_3^{\nu N}\,(x,Q^2)=\frac{Q^2}{ 4\pi^2}
\left[\sigma_{L}(x,Q^{2})-\sigma_{R}(x,Q^{2})\right].
\label{eq:F3}
\end{eqnarray}
where the expression can be interpreted in terms
of parton densities  as being the sea-quark component 
of $xF_3$.  It corresponds to the excitation of the $c\bar s$ state in the process $W^+g\rightarrow c\bar{s}$, with $xF_3$
differing from zero due to the strong left-right asymmetry of the
light-cone $|c\bar s\rangle$ Fock state. For values of Bjorken variable not so small, $xF_3$ contains important valence quark contribution. The valence term, $xq_{val}$, is the same for both  $\nu N$ and
$\bar\nu N$ structure functions of an  iso-scalar nucleon.
The sea-quark ($xq_{sea}$) term in the $xF^{\nu N}_3$ has opposite
sign for $xF^{\bar\nu N}_3$, leading to $xF^{\,\nu (\bar{\nu}) N}_3=xq_{val}\pm xq_{sea}$.

A direct comparison of result in Eq. (\ref{eq:F3}) with experimental data is somewhat difficult. The reason is that structure functions  obtained from neutrino scattering experiments are usually extracted from the sum and the difference of the neutrino and anti-neutrino $y$-dependent differential cross sections, respectively. That is, $xF_3$ is determined by the average $\frac{1}{2}(xF_3^{\nu N}+xF_3^{\bar{\nu} N})$. In Ref. \cite{ZOLLER1}, the authors intend to compare the result of Eq. (\ref{eq:F3}) to CCFR data \cite{CCFR}. However, this procedure is questionable since the $c\bar{s} \,(s\bar{c})$ component disappears in the sum $F_3^{\nu}+F_3^{\bar{\nu}}=2xq_{val}$. This comparison is shown in Fig. \ref{fig:3}, where the structure function $xF_3^{\nu N}(x,Q^2)$ is shown as a function of $Q^2$. The theoretical curves correspond to the results using the GBW (dashed-line) and the IIM (dot-dashed line) models for dipole cross section, respectively.  The total contribution (solid) includes the valence quark contribution through the above procedure. The calculation takes into account nuclear shadowing for the iron nucleus  via Glauber-Gribov formalism, where the dipole-nucleus cross section is computed using Eq. (\ref{sigmanuc}). The distinct models for the dipole cross section give similar results for the range of virtuality considered in the plots. We have checked our results with those in Ref. \cite{ZOLLER1} and it was found they are similar despite the different dipole cross section used. The main reason is that current experimental results are probing mostly the color transparency domain in the dipole cross sections. Future measurements at smaller $x$ and heavy nuclei, as proposed in the Minerva experiment \cite{minerva}, will allow to disentangle the models concerning the parton saturation effects.
%%%%%%%%%%%%%%%%%%%%%%%%%%%%%%%%%%%%%%%%%%%%%%%%%%%%%%%%%%%%%%%%%%%%%%%%%
\begin{figure}[t]
\centerline{\includegraphics[scale=0.5]{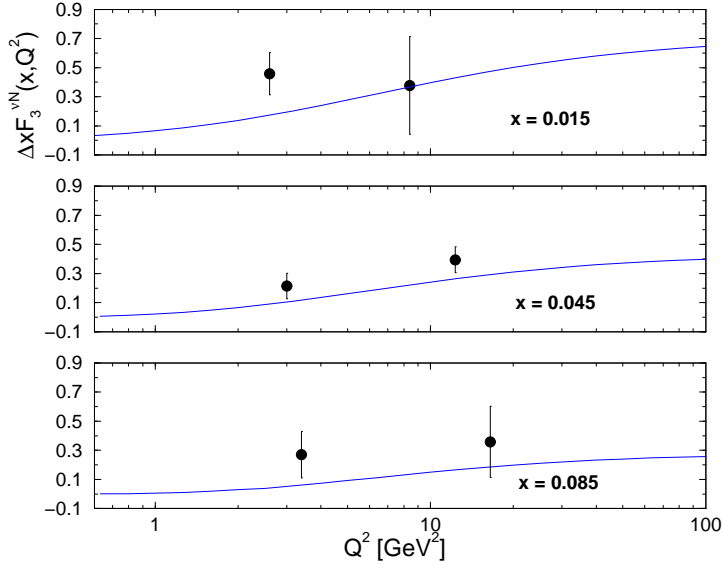}}
\caption{The structure function $\Delta xF_3^{\nu N}$ as a function of boson virtuality (see text).}
\label{fig:4}
\end{figure}
%%%%%%%%%%%%%%%%%%%%%%%%%%%%%%%%%%%%%%%%%%%%%%%%%%%%%%%%%%%%%%%%%%%%%%%%%

The neutrino-antineutrino difference $xF_3^{\nu}-xF_3^{\bar{\nu}}$ provides a determination of the sea (strange) density. In the parton model, one has $xF^{\nu N}_3=xq_{val} -2x\bar{c}(x)+2xs(x)$ and $xF^{\bar{\nu} N}_3=xq_{val}+ 2xc(x)-2x\bar{s}(x)$. Therefore, the neutrino-antineutrino difference effectively measures the strange density, since the charm contribution is small in the kinematical region measured by current experiments. Assuming $s(x)=\bar{s}(x)$ and $c(x)=\bar{c}(x)$ one obtains,
\begin{eqnarray}
\Delta x F_3=xF_3^{\nu N}-xF_3^{\bar \nu N}=2xq_{sea}= 4x\left[s(x)-c(x)\right].\label{eq:deltaf3}
\end{eqnarray}

Here, some comments are in order. Taking into account that our calculation in Eq. (\ref{eq:F3}) corresponds to sea-quark content of $xF_3$, then $\Delta x F_3=2xF_3^{\nu N}$. Our calculation is equivalent to the $cs$ component of the structure function given by the $W$-gluon fusion term at order $\alpha_s$, which reads as,
\begin{eqnarray}
F_3^{\nu N}(W^+g\rightarrow c\bar{s})=\left(\frac{\alpha_s}{2\pi}\right)\int_{ax}^1\frac{dz}{z}g(z,\mu^2)\,C_3\left(\frac{x}{z},\,Q^2\right)\,,
\end{eqnarray}
where $a=1+(m_c^2+m_s^2)/Q^2$ and the Wilson coefficient $C_3$ represents the $W^+g\rightarrow c\bar{s}$ cross section. The total $cs$ contribution to $F_3$ is the sum of the quark excitation term, taken at the factorization scale $\mu^2=m_c^2$ (near threshold), and the gluon-fusion term above. Namely, $F_3^{\nu N}\,(x,Q^2)=2\left[\bar{s}(x_c,\mu^2) - c(x,\mu^2)\right]+ F_3^{\nu N}(W^+g\rightarrow c\bar{s})$,
with the slow-rescaling variable $x_c=x[1+(m_c/Q^2)]$ and similar expression for $F_3^{\bar{\nu} N}\,(x,Q^2)$.

In Fig. \ref{fig:4} the quantity  $\Delta x F_3$ as a function of $Q^2$ at fixed $x$ is shown in comparison with the CCFR result obtained from $\nu_{\mu}Fe$ and $\bar\nu_{\mu}Fe$ differential cross section \cite{CCFR3}.  The theoretical curve is obtained from Eq. (\ref{eq:deltaf3}) using the IIM dipole cross section and Glauber-Gribov shadowing corrections. The agreement is good and the quark excitation contribution, described above, has not been added. This additional piece should improve the description.

As a final analysis, we consider the nuclear ratios for the neutrino structure functions. We are confident in the reliability of these  calculations using the color dipole approach as they were successfully tested against charged lepton scattering data \cite{ARMESTO}. First, we compute the ratio $R_2(x,Q^2)=F_2^{\nu A}/(AF_2^{\nu N})$. This is shown in Fig. \ref{fig:5}-a, as a function of $Q^2$ for fixed $x$ ($10^{-5}\leq x\leq 10^{-2}$). For our purpose, an iron nuclei ($A=56$)  is considered and valence content is disregarded. A strong $Q^2$ dependence is observed. For intermediate virtualities, the ratio ranges on $0.75\leq R_2\leq 0.85$ and decreases with lower $x$. In the current region of neutrino data, the nuclear correction is of order 15-20 \%. This dependence on virtuality is different of the observed flat behavior present in calculation using the DGLAP collinear approach \cite{duancg}. In Fig. \ref{fig:5}-b, it is shown the nuclear ratio $R_3(x,Q^2)=xF_3^{\nu A}/(AxF_3^{\nu N})$ as a function at $Q^2$ at fixed $x$. This corresponds to nuclear shadowing in sea content of $xF_3$. A weak dependence on $Q^2$ is observed, whereas it decreases at lower $x$. An interesting point it is the different shadowing contribution for $xF_3$ in contrast with $F_2$. This feature has already been  observed in the comparison of collinear QCD approach with neutrino data \cite{duancg} and corroborates the dynamical shadowing correction considered here.

%%%%%%%%%%%%%%%%%%%%%%%%%%%%%%%%%%%%%%%%%%%%%%%%%%%%%%%%%%%%%%%%%%%%%%%%%
\begin{figure}[t]
\begin{tabular}{cc}
\includegraphics[scale=0.4]{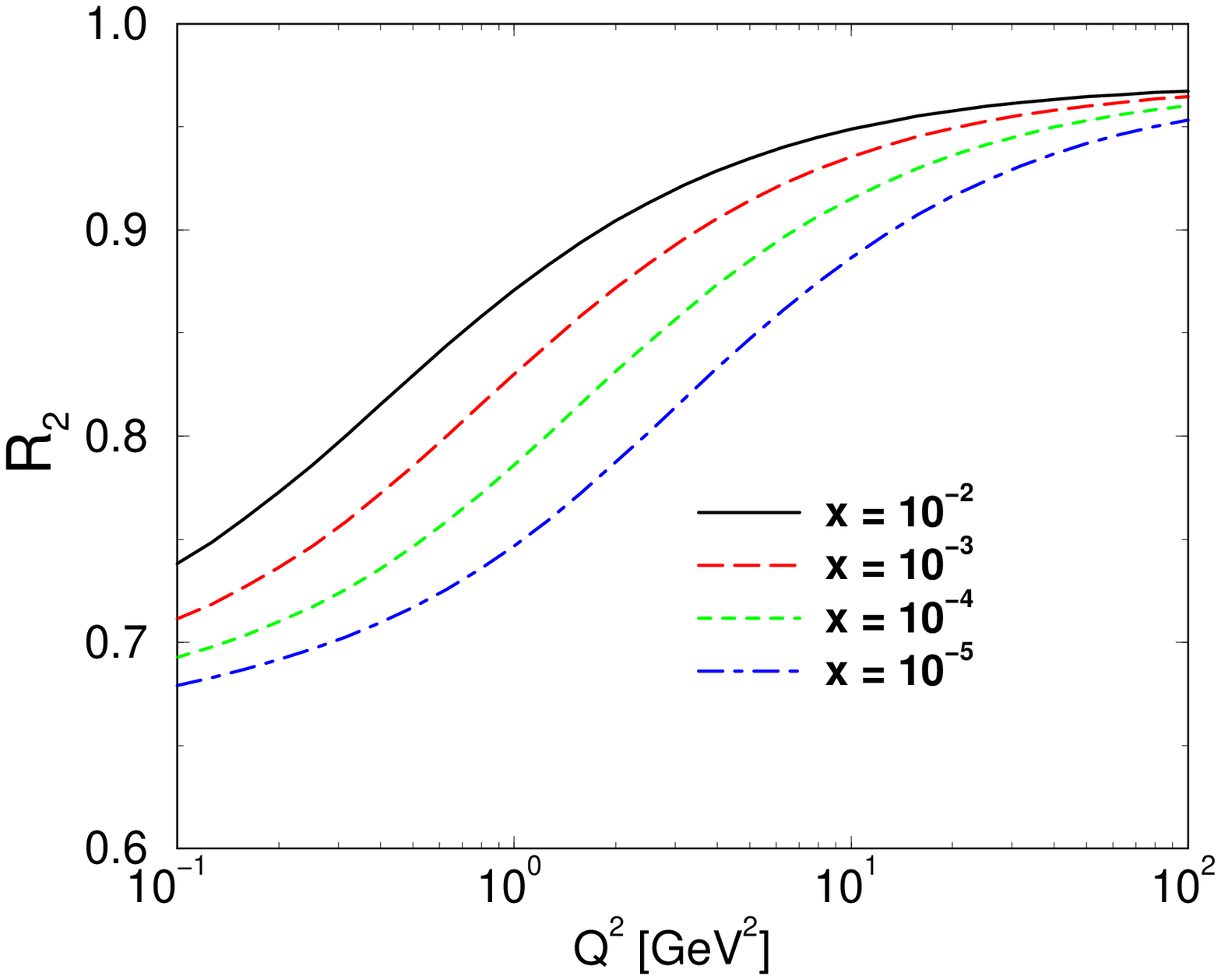} & \includegraphics[scale=0.4]{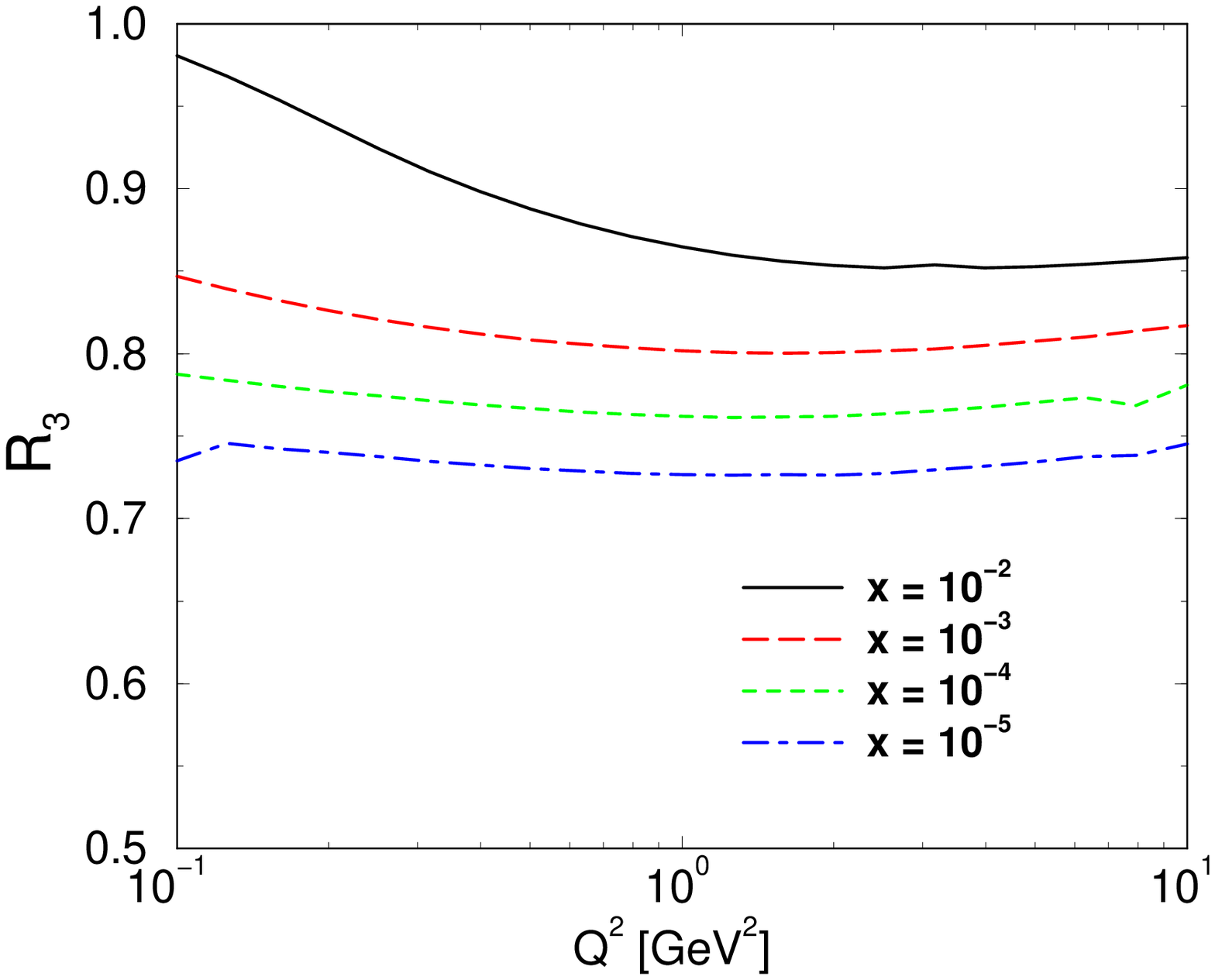} \\
(a) & (b)
\end{tabular}
\caption{The nuclear ratios (a)  $R_2=F_2^A/(AF_2)$  and (b) $R_3=xF_3^A/(AxF_3)$ as a function of virtuality $Q^2$ at fixed $x$ (see text). }
\label{fig:5}
\end{figure}
%%%%%%%%%%%%%%%%%%%%%%%%%%%%%%%%%%%%%%%%%%%%%%%%%%%%%%%%%%%%%%%%%%%%%%%%%

\section{Comments and Conclusions}

As a summary, an analysis of small-$x$ neutrino-nucleus DIS is performed within the color dipole formalism. The structure functions $F_2^{\nu N}$,  $xF_3^{\nu N}$ and the quantity $\Delta xF_3^{\nu N}$ are calculated and compared with the experimental data from CCFR and NuTeV by employing phenomenological parameterizations for the dipole cross section which successfully describe small-$x$ inclusive and diffractive $ep$ DIS data. Nuclear shadowing is taking into account through Glauber-Gribov formalism. It is found that small-$x$ data show geometric scaling property for the boson-hadron cross section as a function of the scaling variable $\tau$. The structure function $F_2$ is in agreement with the phenomenological implementation using the saturation models at the small-$x$ region.  The structure functions $xF_3^{\nu N}$ is also discussed in detail. The sea content, described by the quantity $\Delta xF_3^{\nu N}$, is well described and the addition of quark excitation term should improve it. Although the results presented here are compelling, further investigations are requested. In particular, measurements of neutrino-nucleus structure function in smaller values of $x$ than the currently measured in the accelerator experiments. The present results also confirms the robustness of the  color dipole formalism to describe the total neutrino cross section of ultra-high energy neutrinos. Finally, we predict the nuclear ratios $R_2$ and $R_3$ and single out the size of nuclear effects in each case.

\section*{Acknowledgments}

We thank Donna Naples and U. K. Yang for providing us with the experimental datasets of CCFR Coll. measurements on neutrino-nucleus structure functions. This work is supported by CNPq (Brazil).

\ed